# What happened to modern physics?
## Paul Shabajee and Keith Postlethwaite

20th Century physics appears to tell us that our world is relativistic, quantum mechanical and chaotic. Why do we teach children that it is not?

## Abstract

Relativity, Quantum Mechanics and Chaos theory are three of the most significant scientific advances of the 20th Century - each fundamentally changing our understanding of the physical universe. The authors ask why the UK National Curriculum in science almost entirely ignores them. Children and young people regularly come into contact with the language, concepts and implications of these theories through the media and through new technologies, and they are the basis of many contemporary scientific and technological developments. There is surely, therefore, an urgent need to include the concepts of '20th Century physics' within the curriculum.

## The Problem

One of the major theories in the physical sciences in the 20th Century suggests that the physical world is relativistic, in the sense that measurements of time, space and mass depend on the relative motion of the object and the observer. This theory challenges our expectations which are derived from experiences of everyday objects which normally move at speeds much below the speed of light. However, it still leaves us with a deterministic view of the physical universe in which variables are related in such a way that a change of one variable produces a definite and predictable change in another dependent variable.

Another 20th Century theory suggests that the world is quantum mechanical. This theory introduces many counter-intuitive ideas to predict the behavior of the physical world. For example, the Heisenberg Uncertainty Principle which confronts us with the notion that some properties of particles (e.g. time and energy, or momentum and position) are linked in ways that make it impossible to know both with perfect precision at the same time. In addition to mapping these counter-intuitive behaviours of the small scale universe, quantum theory undermines the concept of determinism through the central role which it gives to probability. For example, the square of the amplitude of a particle's wave function at a given point is the probability of that particle being at that point. What is particularly important here is that probability is not seen to be the consequence of our lack of knowledge about the system, not something we could hope to write out of the model by more complex calculations with more complete data, rather it is fundamental to the system.

The third example of a massive change of view in modern physics is not as well established as relativity or quantum theory. Nevertheless it has overturned the way in which we might seek to model complex behaviors in the physical world, in part by asserting that tiny differences in initial conditions can produce massive differences in outcome, even in systems described by fairly simple and entirely deterministic equations. This is the theory of chaos (Gleick 1988). Chaos theory provides insights into turbulent fluid flow, the behaviour of weather patterns, the behaviour of traffic, the fluctuations of animal populations, and the behaviour of the heart.

It follows that at a fundamental level, Newtonian mechanics, Cartesian geometry and linear causality fail to describe this world adequately even though they still serve us well as working models for many everyday experiences. However, the current UK National Curriculum in Science 5-16 contains no reference to the physical theories which *do* offer at least some appropriate descriptions: relativity, quantum mechanics, statistical processes and chaos theory. Furthermore, in a recent edition of this journal covering the current National Curriculum Review, although Campbell (1998) and Millar et al (1998) propose significant, interesting and largely welcome changes to the pedagogical approach to science education, there was still no mention of introducing 'modern' physical theories into the curriculum.

At higher levels, the situation is somewhat different in that some aspects of quantum behavior get attention in the AS and A2 subject criteria (QCA 1999), some A2 courses contain aspects of modern cosmology, and there are some A2 options which address, for example, special relativity (eg the AQA option, 'Turning Points in Physics' (http://www.aeb.org.uk/)). However, even at A-level, 20th Century physics gets only limited attention and, for example, chaos theory is still largely overlooked. Even the AQA AS syllabus 'Science for Public Understanding' (which might be expected to pick up issues which are prominent in the general media's reporting of science) fails to connect significantly with many of the revolutionary aspects of 20th century physics. This situation at A-level needs (and will undoubtedly get) further attention. However, it is the lack of 20$^{th}$ Century physics in the pre-16 curriculum that is our main concern.

## Why is this a problem?

The new National Curriculum draws attention to a range of purposes for science education (e.g. understanding the use of experimental evidence) which can perfectly well be achieved through study of pre-20th century science. However, it also argues that:

> *"Science stimulates and excites pupils' curiosity about phenomena and events in the world around them. It also satisfies this curiosity with knowledge. Because science links direct practical experience with ideas, it can engage learners at many levels. Scientific method is about developing and evaluating explanations through experimental evidence and modelling. This is a spur to critical and creative thought. Through science, pupils understand how major scientific ideas contribute to technological change - impacting on industry, business and medicine and improving quality of life. Pupils recognise the cultural significance of science and trace its worldwide development. They learn to question and discuss science-based issues that may affect their own lives, the direction of society and the future of the world."*

We feel that these aspects of the revised curriculum are significantly undermined by the absence of attention to 20$^{th}$ Century physics. We argue that there are many reasons why this situation is problematic; here we briefly discuss four of the most fundamental.

Page 2 of 8

**1) Omission of 20th Century physics from the pre-16 curriculum has a negative impact on subsequent learning:**

There is a great deal of research evidence that once children assimilate a mistaken or limited understanding of their world into their cognitive framework, it is difficult to change it (Driver 1985, 1994 ). Hence, for example, by providing children with a 'cognitive framework' derived from Newtonian mechanics and presenting this both as unproblematic and all-powerful (when we know that although it is an excellent working model for predicting 'normal' experiences, it is nevertheless fundamentally conceptually incorrect), we are taking steps designed to make any subsequent learning of 20th Century relativistic ideas extremely difficult.

Another important instance is found in the examples used in lessons, which imply that the world works in an "A leads to B, which leads to C" fashion, most often described by linear equations. Chaos theory indicates the contrary. Indeed only in very special and controlled circumstances, is simple linear causality the case in the physical world.

A cognitive model based on these elements of Newtonian mechanics and simple linear causality will not prepare pupils to embrace modern physics. Little wonder then, that if 'modern physics' is encountered in a later stage of education it is viewed as obscure and impenetrable. Stannard (1999) argues this case strongly:

> *"Our first introduction to the mind-blowing world of modern physics should be when we are young, when our minds are still open and flexible. As we become older we become set in our thinking, our view of the world fossilizes; we become resistant to new modes of thought"*

**2) It reduces the likelihood that the school curriculum will excite pupils:**

In the context of a search for ways in which science education can foster a sense of wonder, enthusiasm and interest in science it seems to us to be a missed opportunity of tragic proportions to deny pupils access to those ideas in physics which are overflowing with intrigue, mystery, beauty and bizarre complexity. Surely carefully guided access to such ideas could stimulate a lifelong excitement about science. Stannard (1999) again argues,

> *"It (i.e. relativity) is absolutely wonderful, amazing science.... But I was angry that no one had told me about it before (undergraduate level). It seemed a scandal that Einstein's theories had been around for so long but were not part of everyday culture."*

Our general point here is also made with regard to 'A' level students by Ogborn (1998) when he says " It cannot be right that much of what appears in popular science books and television programmes is at best marginal in 16-19 courses".

**3) It reduces the perceived relevance of school physics**:

Science in general, and 20th Century physics in particular, are important as elements in the culture to which education seeks to give pupils access. We have a duty to help pupils value science for its cultural importance. The present curriculum denies us the opportunities to explore the *cultural* relevance of modern physics.

Relevance can *also* be seen in terms of the links between science and technology. Certainly it can be argued that much theoretical physics does not have direct technological application and that the people who work within it are not driven by an interest in technology. However, key aspects of 20$^{th}$ Century physics do relate to the technological innovations of our century. We argue that it is this relationship which is likely, initially, to help pupils recognise the relevance of science.

As we move 'Beyond 2000', the technologies of TV, CD ROM and the Internet regularly expose children to talk of galaxies, black holes, alternative universes, hyper-space and theories of time travel (see, for example, the 1999 Royal Institution Lectures, details on the BBC website - see further reading). Children use computer systems whose semiconductor- based processors and CD-ROM lasers, harness quantum mechanical effects. Newer technologies are reported on popular science TV programmes using terms like 'super conductors', 'positron emission tomography scanners', 'quantum computers and cryptography' and 'relativity', the last, for example in relation to 'atomic clocks' in 'global positioning system satellites'. There is a startling contrast between the physics that is relevant to these technologies and the physics that children are taught in schools, which deals predominantly with batteries and bulbs, trolleys on ramps, and weightless strings and frictionless pulleys. We argue that it is important that the science that relates to the discovery and understanding of 20$^{th}$ Century technologies should described, discussed and debated with pupils.

Some seek to weaken this argument by suggesting that 20$^{th}$ Century technologies *might* have been developed without input from physicists working in 'pure science'. However, in the majority of cases they were not: for example transistors were invented by people who "were versed in and contributed to the theory of solids", and computer circuits were designed by people "dealing with the counting of nuclear particles" (Hey & Walters, 1987 p vi).

These three issues may make school physics less engaging, and the prospect of moving on to radically different, and as yet unmentioned 'modern' ideas, quite daunting. In both these ways the present curriculum might contribute to the difficulties which physics courses in higher education face in recruiting students. Our final reason for questioning the omission of 20th century physics is of a somewhat different character:

**4) It limits our pedagogical imagination:**
If you spend an evening at home with many primary school age children you are likely to see that concepts of hyper-space and time travel, alternative/multiple realities and the implications of chaos theory are well with in their grasp, as they see, talk about and explore TV programmes, and computer games. However, the relative 'safety' of the National Curriculum does little to encourage teachers to explore ways to deepen pupils' understanding of these and other elements of modern physics. The present authors certainly do not claim to have the answers to how this might best be done, though it is clear from authors such as Feynman that answers can be found (Feynman 1988).

Feynman diagrams express the complex physics of Quantum Electrodynamics through 'visual' models which are simple to understand and conceptually accurate, and do not have to be unlearned in order for the learner to add the more 'difficult' mathematical models to her or his understanding at a later stage. Visual models like these (which may be particularly suited to animated presentation via ICT) may be one way to get the ideas across. Krauss (1998) suggests a different route: he argues that, with care, it may be productive to use science fiction to teach physics. This is an approach taken by Stannard (1991) who has written about relativity for 11 year olds.

We are excited by a third approach. The 'space' children inhabit is no longer limited by their personal physical experience or their physical human scale. It is through the extension of their nervous systems by technology that they experience 'very small', 'very large and indeed 'entirely fictitious' realities on a daily basis. In this approach, computer graphics technology can again have a key role, enabling pupils (and teachers) to begin to 'visualize' these realities. The 21st century Virtual Reality equivalent of 'Mr Tomkins in Wonderland' (Gamow, 1939) is therefore a real (and exciting) possibility!

The absence of $20^{th}$ Century physics in the curriculum reduces the need for teachers to think creatively about these aspects of innovative pedagogy – aspects which, once explored, might well enhance our teaching of more traditional curriculum areas too.

**The Reasons?**

Why then do we not teach these fundamental physical principles to young children?

Perhaps we feel that children are not able to cope with the ideas of modern physics, but what is the real evidence for this? Perhaps we think that the effects of these theories are not relevant to pupils' lives; but how, then, do we counter the argument that the entertainment world of the school children is full of direct references and the technology they use is dependent on the theories? Perhaps we think that children will learn best if they are taught in the chronological order of theoretical development, but again, what is our evidence, and how do we account for the fact that a truly historical approach to the current curriculum would take us up a variety of backwaters? Perhaps we feel that because the mathematics of modern physics is complicated, the phenomena and concepts themselves cannot be adequately taught. If so Feynman diagrams (see above) provides a startling counter example, demonstrating that the concepts are themselves not 'difficult'. Perhaps we fail to emphasize modern physics because there is not a widespread understanding of the importance of its insights by the public or policy makers? Perhaps we think that the theories are of no practical use. Perhaps we are concerned that we simply do not have the teachers able to teach it?

We believe the answers are likely to be all of the above (and no doubt many more). We also believe that few, if any, of these supposed answers are valid.

A survey of the literature seems to offer little help, as the question we have posed does not seem to have been widely asked, except in papers that seem to start with the premise that learning of these principles starts in early adulthood. We feel that it is time for these issues to be discussed more widely and in depth.  It is too late for any such discussions to make an impact on the recently completed schools curriculum review. Perhaps during the next review, 20th Century physics can take its place in the 21st Century curriculum.

**Moving Forward?**

The above critique it is clearly meant to stimulate discussion and puts forward arguments for the teaching of ideas and concepts of 'modern physics' from an early age.  If we were to assume that in principle this is desirable, how might we move forward?

This is a complex issue, approachable at a number of levels.  Some approaches could be implemented fairly simply.  One such short-term approach (that any of us could use in our own classrooms) is to begin to embed some of the most fundamental lessons of 'new physics', into our day to day teaching, without actually changing the content of what is taught. For example, by carefully replacing language of 'absolute causality' by that of the more probabilistic and complex world view indicated by quantum mechanics and chaos theory - replacing "this *is* what will happen when…' to 'this *is the most likely thing* that that will happen when…"  or  "this is the equation which *describes*…" with "this equation gives us *a good approximation of* …". The potential impact on this relatively simple change could be dramatic - since the more conditional phrasings are both more accurate (and honest) and automatically beg the questions which form the basis of what for many physicists is the joy of physics, that there is always a 'better' model waiting to be discovered... for the moment at least.

There are many other changes that could be made to the curriculum which would 'bring' modern physics more directly into schools. We do not claim to have answers in that regard. Some very fundamental issues arise in that direction and many debates about those need to take place. The issues include: the aims and role of science education in our society, the detailed nature of any such changes, the pressures that any desired change would cause to an already over burdened profession, the potential of children to understand these 'complex' ideas, how teachers could be supported to teach these topics and the political implications of any and all of the above.  What is clear to us is that such debate cannot be replaced by centrally imposed curriculum change.  Teachers, educationalists and physicists need to be engaged, together, in creating new ways of working.

**Conclusion**

We have put forward a range of arguments for the urgent need to consider teaching the ideas and concepts of '20th Century physics' as a fundamental part of school science education across the age range. We are not arguing that this *must* be the case, only that there appear to be some very good reasons for such a stance, that reasons for not doing so appear open to question, and that a debate about such issues is called for.

The urgency is highlighted by the following eloquent quotation from a paper by Eric Rogers (1969) in which he asked whether the content of the physics curriculum was useful or necessary, and offered some specific ways forward with regard to teaching quantum mechanics at school.

> *"In thinking towards the future of our teaching, I do not know how much we can throw out; I do not know how much new teaching we can bring in; and I cannot yet see clearly how we shall present our teaching; but I would make prediction, based on my trust in flexibility of mind. Remember Whitehead's dictum that the aim of education should be to foster and preserve flexibility of mind. If we are flexible in our ideas and look forward to quantum mechanics being a principal part of our teaching, we can go so far, that in the Year 2000, our teaching will be different: it will do justice to modern science. Our young students will say: "I am acquainted with the way physicists think, talk, and speculate about the physical world". That is my hope, but it is for a new generation of teachers to carry it out."*

Sadly, there seems to have been little change in relation to these issues over the last 30 years. Can we become the "new generation of teachers" of whom Rogers spoke? Not it seems, in the year 2000, but perhaps we can begin hotly debating these ideas and questions, and bringing the fundamental lessons of 20th Century physics to the science classrooms of the 21st Century.

**Further Reading**